\documentclass{wscpaperproc}
\usepackage{latexsym}
\usepackage{graphicx}
\usepackage{mathptmx}
\usepackage[T1]{fontenc}

\usepackage{amsmath}
\usepackage{amsfonts}
\usepackage{amssymb}
\usepackage{amsbsy}
\usepackage{amsthm}
\usepackage{booktabs}
\usepackage{multirow}
\usepackage{threeparttable}
\usepackage{caption}
\usepackage{subcaption}

\usepackage[pdftex,colorlinks=true,urlcolor=blue,citecolor=black,anchorcolor=black,linkcolor=black]{hyperref}

\newtheoremstyle{wsc}%
{3pt}%
{3pt}%
{}%
{}%
{\bf}%
{}%
{.5em}%
{}%

\theoremstyle{wsc}

\begin{document}

\pagestyle{fancyplain}

\thispagestyle{plain}
\firstPageHead{}

\chead{\fancyplain{}{\itshape Chen, Chang, Power and Jing}}

\rhead{}
\cfoot{}
\renewcommand{\headrulewidth}{0pt} 

\input{wscbib.tex}           

\setlength{\baselineskip}{12.7pt}

\title{An Integrated Multi-Physics Optimization Framework for Particle Accelerator Design}

\author{Gongxiaohui Chen, Tyler H. Chang, John Power\\[12pt]
	Argonne National Laboratory\\
	9700 S. Cass Ave, Lemont, IL 60516
\and
    Chungunag Jing\\[12pt]
	  Euclid Techlabs LLC\\
	367 Remington Blvd, Bolingbrook, IL 60440
}

\maketitle
\section*{Abstract}

The overarching goal of beamline design is to achieve a high brightness electron beam from the beamline. Traditional beamline design
studies involved separate optimizations of radio-frequency cavities, magnets, and beam dynamics using different codes and pursuing various intermediate objectives. In this work, we present a novel unified global optimization framework that integrates multiple physics modules for beamline design as simulation functions for a two-stage global optimization solver.

\section{Introduction}
\label{sec:intro}

The design of an optimized beamline is a complex and labor-intensive task that involves combining the dynamics of interacting particles with external fields produced by elements such as cavities or magnets. In traditional design studies, the various components, such as beam dynamics, cavities, and magnets, were optimized independently by separate individuals using different codes with isolated targeting objectives. 

Given the ultimate goal of producing high brightness beam through the designed beamline, we propose a unified framework which include three modules: 1) electromagnetic module (EMM) for solving the resonant cavity eigenmodes field (based on \verb|SUPERFISH|), 2) magnetostatic module (MSM) for magnet studies (based on \verb|POISSON| or parameter controlled extrapolated Bz curve), 3) beam dynamics module (BDM) for particle tracking (based on \verb|ASTRA|). A docker image with \verb|SUPERFISH|/\verb|POISSON|~\cite{docker-poisson-superfish-nobin} and a python tool~\cite{PySuperfish} were used through the optimization work. To effectively handle a large number of tuning variables during the global optimization process, a {\sl localized} model-based optimization method was employed, which requires fewer simulation evaluations than other comparable methods.

\section{Global optimization workflow for accelerator design} 
\begin{figure}[h!]
    \centering
    \includegraphics[width=0.92\textwidth]{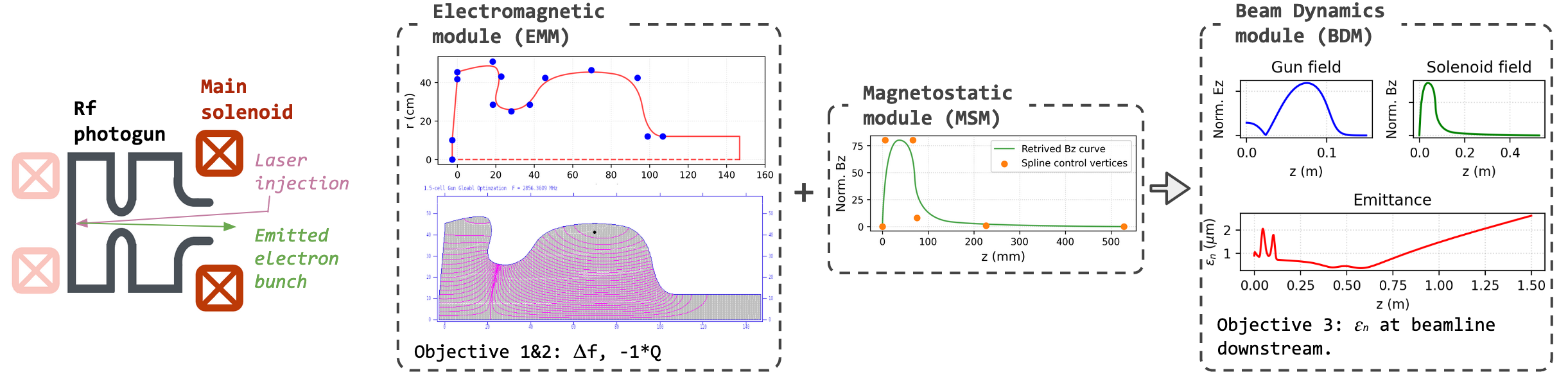}
    \caption{A schematic diagram of a simplified beamline (left), and workflow of a single evaluation (right).}%
    \label{fig:workflow}
\end{figure}

To demonstrate the global optimization workflow, we employed the S-band (2.856 GHz) BNL type 1.5-cell photogun as the prototype. As shown in Figure~\ref{fig:workflow}, coordinates of 14 control vertices were partially used as input variables, offering high flexibility in adjusting the gun geometry in the EMM. The followed main solenoid was fine-tuned by 6 control vertices to achieve an ideal field distribution (featuring a sharp rise at the cathode surface). In practical applications, the MSM had the flexibility to interchange with a real solenoid prototype model, allowing geometry tuning using \verb|POISSON|. Then the on-axis $E_z$ of the photogun from EMM and the on-axis $B_z$ of the main solenoid from MSM were used as external fields in the beam dynamics module (BDM) for conducting beam dynamics simulations (Figure~\ref{fig:workflow}). The tuning variables in the BDM include the gun phase and solenoid peak field.

To perform the global optimization we use a model-based multiobjective solver built with ParMOO \cite{parmoo} and distributed simulation evaluations on the high-performance computing (HPC) system Bebop at Argonne using libEnsemble \shortcite{libensemble}.
In ParMOO, we model each of the 23 input variables across all three modules as a continuous design variable.
The unified framework is modeled as a single ParMOO simulation function, with multiple outputs including $\Delta f$ (defined by $\left|f_{target} - f_{simulate}\right|$, where $f_{target}$ is of 2.856~GHz), -Q (-1 times the quality factor), and transverse beam emittance ($\varepsilon_n$) downstream of the beamline. In order to steer ParMOO away from poor geometries, a few constraints are implemented, which include particle death ratio of <~0.5$\%$, $\Delta f$ of <~0.5~MHz, and maintaining a reasonable field balance between cells to ensure the desired operating $\pi$-mode. 

In order to efficiently tune 23 input variables (high-dimensional by global optimization standards) on a limited budget, we perform a global search phase via a 800 point Latin hypercube design, followed by 300 iterations of localized Gaussian process modeling and trust-region descent.
For further information, a similar method is described on the \href{https://parmoo.readthedocs.io/en/latest/tutorials/local_method.html}{ParMOO docs page}.
In each iteration of trust-region descent, we employed 15 randomized scalarizations to achieve coverage of the entire Pareto front, and fixed 1 scalarization in order to specifically target low emittance solutions.
This results in a batch size of 16, and
a total simulation budget of 5600 over all 300 iterations.

\section{Results and Conclusion}
\label{sec:results}

One of the critical challenges is handling 23 variables while adhering to the strict constraints required by the operation standard. To address this challenge, we employed a model-based trust-region optimizer that efficiently handles a large number of design variables in the optimization process. By a given initial beam source with spot size of 0.5~mm, bunch length (FWHM) of 300~fs, bunch charge of 100~pC, along with a standard operation gun gradient of 150~MV/m, the emittance generated by the optimized beamline was found to converge to approximately 0.3~$\mu$m (one of these optimized results is depicted in Figure~\ref{fig:workflow}), which is comparable to state-of-the-art results. 

The framework offers great flexibility in accelerator design, making it easier to explore various physics concepts and obtain more statistical results through systematic geometry tuning (e.g. changing cell lengths or adding concave features to introduce a focusing field on cathode).

\section*{Acknowledgments}
\footnotesize
This work is supported by Laboratory Directed Research and Development (LDRD) funding from Argonne National Laboratory.
This work was also supported in part by the U.S.~Department of Energy, Office of
Science, Office of Advanced Scientific Computing Research and Office of
High-Energy Physics, Scientific Discovery through Advanced Computing (SciDAC)
Program through the FASTMath Institute and the CAMPA Project under Contract
No.~DE-AC02-06CH11357.
We gratefully acknowledge the computing resources provided on Bebop, a HPC cluster operated by the Laboratory Computing Resource Center at Argonne National Laboratory. 
We thank Philippe Piot for the insightful discussions on the simulations.
We also thank Jeff Larson, John-Luke Navarro, and Steve Hudson for their advice on libEnsemble usage.
\footnotesize

\bibliographystyle{wsc}
\bibliography{refs}

\begin{thebibliography}{}

\bibitem[\protect\citeauthoryear{Chang and Wild}{Chang and Wild}{2023}]{parmoo}
Chang, T.~H., and S.~M. Wild. 2023.
\newblock ``{ParMOO}: A {P}ython library for parallel multiobjective simulation
  optimization''.
\newblock {\em Journal of Open Source Software\/}~8(82):4468.


\bibitem[\protect\citeauthoryear{Hudson, Larson, Navarro, and Wild}{Hudson
  et~al.}{2022}]{libensemble}
Hudson, S., J.~Larson, J.-L. Navarro, and S.~Wild. 2022.
\newblock ``{libEnsemble}: A Library to Coordinate the Concurrent Evaluation of
  Dynamic Ensembles of Calculations''.
\newblock {\em {IEEE} Transactions on Parallel and Distributed
  Systems\/}~33(4):977--988.


\bibitem[\protect\citeauthoryear{Mayes}{Mayes, Christopher}{2023}]{PySuperfish}
Mayes, Christopher 2023.
\newblock ``PySuperfish''.
\newblock Available at: \url{https://github.com/ChristopherMayes/PySuperfish}.

\bibitem[\protect\citeauthoryear{Slepicka}{Slepicka,
  Hugo}{2020}]{docker-poisson-superfish-nobin}
Slepicka, Hugo 2020.
\newblock ``Poisson Superfish via Docker''.
\newblock Available at:
  \url{https://github.com/hhslepicka/docker-poisson-superfish-nobin}.

\end{thebibliography}

\end{document}